\documentclass[12pt,a4paper]{article}
\usepackage{graphicx}
\usepackage{amsmath}
\usepackage{amssymb}
\usepackage{cite}
\usepackage[height=8.8in,width=6.45in]{geometry}

\newcommand{\roubleit}{{\it{Р}\kern-.50em\rule[.3ex]{.38em}{.03em}\kern.07em}}

\begin{document}
\begin{titlepage}
\hfill CALT-TH 2017-014, IPMU 17-0041\\
\vbox{
    \halign{#\hfil         \cr
           } % end of \halign
      }  % end of \vbox
\vspace*{17mm}
\begin{center}
{\Large \bf New Kaluza-Klein Instantons and Decay of AdS Vacua}

\vspace*{15mm}

{\large Hirosi Ooguri${}^{1, 2}$ and
 Lev Spodyneiko${}^{1, 3}$}

\vspace*{8mm}

${}^1$ Walter Burke Institute for Theoretical Physics \\ California
Institute of Technology,
 Pasadena, CA 91125, USA\\
\vspace*{0.3cm}
${}^2$ Kavli Institute for the Physics and Mathematics of the Universe
\\ University of Tokyo,
Kashiwa, 277-8583, Japan\\
\vspace*{0.3cm}
${}^3$
Landau Institute for Theoretical Physics, 142432 Chernogolovka, Russia

\vspace*{9mm}

%%\maketitle

\end{center}
\begin{abstract}
We construct a generalization of Witten's Kaluza-Klein instanton, where a
higher-dimensional sphere (rather than a circle as in Witten's instanton) collapses
to zero size and the
geometry terminates at a bubble of nothing, in a low energy effective theory of M theory.
We use the solution to exhibit instability of non-supersymmetric AdS$_5$ vacua
in M Theory compactified on positive K\"ahler-Einstein spaces,
providing a further evidence for the recent conjecture that
  any non-supersymmetric anti-de Sitter vacuum supported by fluxes must be unstable.

\end{abstract}

\end{titlepage}

\vskip 1cm

\section{Introduction}

Stability is an important criterion
for consistency of Kaluza-Klein vacua.
Due to non-trivial topologies of their internal spaces,
the standard positive energy theorem \cite{Schon:1979rg, Schon:1981vd,
Witten:1981mf}
does not necessarily apply.
In fact, it was shown by E.~Witten \cite{Witten:1981gj}  that
the original Kaluza-Klein theory \cite{Kaluza:1921tu, Klein:1926tv}
on a product of the four-dimensional Minkowski spacetime and a circle
is unstable against  a semi-classical decay process, unless protected by
boundary conditions on fermions.
The instanton that mediates the
decay is the analytical continuation of five-dimensional Schwarzschild solution,
\begin{align*}
ds^2=\frac{dr^2}{1-\frac {R^2}{r^2}} +r^2 d\Omega_3^2+\left(1-\frac{R^2}{r^2}\right)d\phi^2,
\end{align*}
where $d\Omega_3^2$ is metrics on the unit three-shpere, and $\phi$
is the coordinate on the Kaluza-Klein circle.
Smoothness of the solution at $r=R$ requires $\phi$ to be periodic
with period $2\pi R$, and the Kaluza-Klein radius at $r=\infty$ is $R$.
As we move toward small $r$, the circle collapses and becomes zero size at $r=R$, where the geometry terminates.
Another analytic continuation of the polar angle on the three-sphere
turns this into a Lorentzian signature solution, where
the ``bubble of nothing'' expands with velocity that asymptotes to the speed
of light.

Witten's instanton has also played a role in stability of anti-de Sitter (AdS) vacua.
There have been several proposals for non-supersymmetric AdS geometries.
Among them is AdS$_5 \times S^5/\Gamma$, where $\Gamma$ is
a discrete subgroup of the $SU(4)$ rotational symmetry of $S^5$ \cite{Kachru:1998ys}.
Supersymmetry is completely broken if $\Gamma$ does not fit
within an $SU(3)$ subgroup of the $SU(4)$ symmetry.
It turns out that, if $\Gamma$ has
a fixed point on $S^5$ or if the radius of $S^5$ is not large enough, the perturbative spectrum on AdS$_5$ contains closed string tachyons
that violate the Breitenloner-Freedman bound
\cite{Dymarsky:2005uh, Dymarsky:2005nc}.
When $\Gamma$ has no fixed point
and  $S^5$ is sufficiently large, the instability modes are lifted and
the configuration becomes perturbatively stable. However, in this case,
$S^5/\Gamma$ is not simply connected, and
 there is a Witten-type instanton where
a homotopically non-trivial cycle on $S^5/\Gamma$ collapses to
zero size at a bubble of nothing~\cite{Horowitz:2007pr}.
This eliminates AdS$_5 \times S^5/\Gamma$ as a candidate for
a stable non-supersymmetric AdS geometry.

In this paper, we present instanton solutions where a higher-dimensional sphere rather than a circle collapses, in a low energy
effective theory of M theory. Such
generalizations of Witten's instanton were
attempted earlier, for example in  \cite{Young:1984jv}, where it was found
that fluxes
needed to cancel the intrinsic curvature on the sphere prevent it from collapsing.
Motivated by the recent conjecture \cite{Ooguri:2016pdq} (see also \cite{Freivogel:2016qwc, Danielsson:2016mtx, Banks:2016xpo})
that  any non-supersymmetric AdS
vacuum supported by fluxes must be unstable, we found examples in non-supersymmetric setups which avoid the
difficulty in the earlier attempt.

We will focus on AdS$_5$ times positive K\"ahler-Einstein spaces, which break supersymmetry \cite{Pope:1988xj}.
AdS$_5 \times \mathbb{C}{\rm P}^3$ is the only example of this type known to be stable against linearized supergravity perturbations\cite{Martin:2008pf}. Since its internal space is simply connected,
 we do not expect
it to have a Witten-type instanton, where
an $S^1$ collapses to zero size at a bubble. On the other hand,
$\mathbb{C}{\rm P}^3$ can be realized as $S^2$ fibration over $S^4$, and it is possible
for its $S^2$ fibers to collapse.
Indeed, we find such instanton solution with finite action.

Our solution avoids the difficulty with fluxes encountered in \cite{Young:1984jv} as follows.
The AdS$_5$ geometry in question is supported by the 4-form flux in M Theory,
with non-zero components both on the $S^2$ fibers and on the $S^4$ base
of the internal space.
As we move toward the center of AdS$_5$, the 4-form flux re-orient itself. By the time we reach the bubble of nothing,
the flux has no components in the $S^2$ fiber direction. Thus, the $S^2$ can collapse at the bubble without violating the
flux conservation.

AdS$_5 \times \mathbb{C}{\rm P}^3$  is only marginally stable
with normalizable mode at the Breitenlohner-Freedman bound. Thus, this vacuum is also in danger of becoming unstable
by higher derivative corrections to the eleven-dimensional supergravity.
It is interesting to point out that in our instanton solution
 the normalizable mode at the Breitenlohner-Freedman bound is turned on
and is responsible for triggering the collapse of the $S^2$ fiber.

If there is a ``bubble of nothing'' instanton in AdS,
it causes an instability that
can
be detected instantaneously on the boundary of AdS \cite{Horowitz:2007pr, Harlow:2010az}. This is because
any observer in AdS can receive signals from any point on a Cauchy
surface within a finite amount of time, and an observer at the
boundary
in particular has access to an infinite volume space near the boundary within an infinitesimal
amount of time. Therefore,
our new instanton solution in
the perturbatively stable non-supersymmetric AdS$_5$ configuration
offer further evidences for the conjecture of  \cite{Ooguri:2016pdq}.

The plan of the paper is the following. In section~\ref{ads cp} we describe the AdS$_5\times \mathbb{C}{\rm P}^3$  solution
and introduce
our instanton ansatz. Boundary conditions on the instanton at the bubble and
at the infinity
are discussed in section~\ref{sec asym}. It turns out that there are algebraic
relations among variables in our ansatz, as shown in section~\ref{anal rel}.
These relations reduce the problem to a second order ordinary differential equation
on a single function, which we will numerically solve in section~\ref{num sol}.
Finiteness of the instanton action is verified in section ~\ref{inst action}.
In section ~\ref{cp1 cp2 inst}, we discuss
 AdS$_5\times \frac {SU(3)}{U(1)\times U(1)}$. It is not know whether this geometry
 is stable against linearized supergravity perturbation. Regardless, we will show that
 it allows a bubble of nothing solution and is therefore unstable non-perturbatively.
In the final section, we discuss additional features of our
instanton solutions.

\section{ AdS$_{\bf 5} \times \mathbb{C}{\rm P}^{\bf 3}$ Geometry and Instanton Ansatz} \label{ads cp}

For any K\"{a}hler-Einstein sixfold $M_6$, there exists an AdS$_5\times M_6$
solution \cite{Pope:1988xj} to  the eleven-dimensional supergravity equations of motion,
\begin{align}
\begin{split}\label{sugra eq}
R_{MN}&=\frac 1 {3} (F_{M}{}^{PQR}F_{NPQR}-\frac 1 {12}g_{MN}F^{PQRS}F_{PQRS}),
 \\
\nabla_{M} F^{MPQR}&=-\frac 1 {576} \varepsilon^{M_1\dots M_8 PQR}F_{M_1M_2M_3M_4}F_{M_5M_6M_7M_8}.
\end{split}
\end{align}
Such a solution
can be found by setting the 4-form field strength as,
\begin{align}\label{F ansatz}
F= c\, \omega\wedge \omega,
\end{align}
where $\omega$ is the K\"{a}hler 2-form of internal space and $c$ is some constant, which will be related to the AdS radius.
With this ansatz, the right-hand side of
the second equation in  (\ref{sugra eq})
vanishes since $F$ is nonzero only on $M_6$, and
the left-hand side vanishes by the
K\"{a}hler integrability condition on $\omega$.
On the other hand, the first equation in  (\ref{sugra eq}) gives,
\begin{align}
R_{\mu\nu}=- 2 c^2 g_{\mu\nu}, \quad
R_{mn}=2c^2 g_{mn},
\end{align}
where $\mu=0,\dots, 4$ is index in the non-compact directions
and $m=5,\dots, 10$ are on $M_6$. Therefore, the non-compact directions  can be chosen to be AdS$_5$, and $M_6$ must be
an Einstein manifold.
The configuration breaks supersymmetry \cite{Pope:1988xj,Gauntlett:2004zh}
since there are no non-trivial solutions to $\delta \Psi_M=0$ for
supersymmetry variation of the gravitini,
\begin{align}
\delta \Psi_M = D_M \varepsilon = \nabla_M \varepsilon+\frac 1 {144} \big(\Gamma_{MNPQR}F^{NPQR}-8\,\Gamma_{NPQ} F_{M}{}^{NPQ}\big)\,\varepsilon .
\end{align}
As we mentioned in the introduction, the only known perturbatively stable
case is $M_6 = \mathbb{C}{\rm P}^3$.
This space can be realized as $S^2$ fibration, and we look
for instanton solution where the fiber collapses. In the next few sections
we focus on $M_6 = \mathbb{C}{\rm P}^3$ and in section \ref{cp1 cp2 inst} we show that instanton solution for $M_6 =\frac{SU(3)}{U(1) \times U(1)}$ can be constructed similarly.

To make the $S^2$ fibration explicit, we use the following set of coordinates
on $\mathbb{C}{\rm P}^3$ \cite{Duff:1986hr}:
\begin{align}\label{vier cp3}
\begin{split}
e_1&=\sqrt{g(r)}d\mu, \\
e_i&=\frac {\sqrt{g(r)}} 2\sin\mu \,\Sigma_{i-1} \quad \text{for } i=2,3,4,\\
e_5&= \sqrt{h(r)} (d\theta - A_1\sin \phi +A_2\cos \phi ), \\
e_6&=\sqrt{h(r)}\sin \theta(d\phi - \cot\theta(A_1\cos\phi +A_2\sin\phi )+A_3),
\end{split}
\end{align}
where
\begin{align}
\begin{split}
\Sigma_1&= \cos\gamma\,d\alpha+\sin\gamma \,\sin \alpha \,d\beta,\\
\Sigma_2&= -\sin\gamma\,d\alpha+\cos\gamma \,\sin \alpha \,d\beta,\\
\Sigma_3&= d\gamma+\cos\alpha \, d\beta,\\
A_i&=\cos\left(\frac \mu 2\right)^2 \Sigma_i.
\end{split}
\end{align}
Here the first 4 tetrad corresponds to the base $S^4$ and the last two correspond to the $S^2$ fiber. We multiplied them by the functions $g(r)$, $h(r)$ to make their sizes dynamical. We take the vierbein on Euclidean AdS space to be,
\begin{align}
\begin{split}
e_7&=dr,\\
e_k&=\sqrt{f(r)}\, \widehat e_k \quad \text{for } k=8,9,10,11,
\end{split}
\end{align}
where $\widehat e_k$ is any tetrad on the $S^4$.
The metric in this frame is,
\begin{align}\label{metrics}
\begin{split}
ds^2=g(r) \left(d\mu^2+\frac 1 4 \sin^2\mu\, \sum_{i=1}^3 \Sigma^2_i\right)+h(r) \Big(d\theta - A_1\sin \phi +A_2\cos \phi \Big)^2 \\+h(r)\sin^2 \theta\Big(d\phi - \cot\theta(A_1\cos\phi +A_2\sin\phi )+A_3\Big)^2+dr^2+f(r)d\Omega_4^2.
\end{split}
\end{align}
We used the freedom of coordinate redefinition by fixing the coefficient near $dr^2$ to be 1.

The next step is to write an ansatz for the 4-from field, utilizing the $SU(3)$-structure of the squashed $\mathbb{C}{\rm P}^3$ given by 2-form $J$ and 3-form $\Omega$
as \cite{Tomasiello:2007eq,Koerber:2008rx,Aldazabal:2007sn},
\begin{align}\label{su3 def}
\begin{split}
J&=-\sin\theta \cos \phi (e^{12}+e^{34})-\sin\theta\sin\phi (e^{13}+e^{42})-\cos\theta (e^{14}+e^{23})+e^{56},\\
\text{Re}\ \Omega&= \cos\theta \cos\phi (e^{126}+e^{346})+\cos\theta\sin \phi (e^{136}+e^{426})+\sin\phi (e^{125}+e^{345})\\&-\cos\phi (e^{135}+e^{425})-\sin\theta (e^{146}+e^{236}),\\
\text{Im}\ \Omega&= -\cos\theta \cos\phi (e^{125}+e^{345})-\cos\theta\sin \phi (e^{135}+e^{425})+\sin\phi (e^{126}+e^{346})\\&-\cos\phi (e^{136}+e^{426})+\sin\theta (e^{145}+e^{235}).
\end{split}
\end{align}
Here $e^{12}=e^1\wedge e^2$ e.t.c..
These forms satisfy,
\begin{align}
\begin{split}\label{su3 str}
d_6J&=\frac 3 2 W_1 \ \text{Im}\ \Omega,\\
d_6\text{Im}\ \Omega&=0,\\
d_6 \text{Re}\ \Omega&=W_1\ J \wedge J+W_2\wedge J,
\end{split}
\end{align}
where $d_6$ is external derivative of $\mathbb{C}{\rm P}^3$, and $W_1$ and $W_2$ are torsion classes of the $SU(3)$-structure
given by,
\begin{align}\label{torsion}
\begin{split}
W_1& = \frac 2 3 \frac{g(r)+h(r)}{g(r)\sqrt{h(r)},}\\
W_2&=  \frac{2h(r)-g(r)}{g(r)\sqrt{h(r)}}\left(\frac 2 3 J-2 e^{56}\right).
\end{split}
\end{align} A general manifold with $SU(3)$ structure has more terms in the relations~(\ref{su3 str}), but in the case of our interest (squashed $\mathbb{C}{\rm P}^3$) other torsion classes vanish.

Note, that this $SU(3)$-structure is different from the usual Fubini-Study K\"{a}hler structure of $\mathbb{C}{\rm P}^3$. We use $\omega$ to
denote the Fubini-Study K\"{a}hler 2-from to distinguish it from $J$. These two $SU(3)$-structures are associated to
different realizations of $\mathbb{C}{\rm P}^3$ as coset spaces. The first is $\mathbb{C}{\rm P}^3=\frac{SU(4)}{U(3)}$, which is a symmetric space and the complex structure of $SU(4)$ gives the Fubini-Study structure. The second is $\mathbb{C}{\rm P}^3=\frac {Sp(2)}{S(U(2)\times U(1))}$, which
is not manifestly symmetric but homogeneous. Therefore, we can use the latter even after we change the relative sizes of the base $S^4$ and the fiber $S^2$. This is the reason why we use the second structure to build an ansatz for the 4-form field.

Left-invariant 2-forms and 3-forms are
spanned by $J,W_2$ and $\text{Re}\ \Omega,\text{Im}\ \Omega$ respectively. Therefore, the most general ansatz for the 4-form respecting the symmetries is
\begin{align}\label{fourforms}
F_4 = \xi_1(r) J\wedge J + \xi_2(r) J\wedge e^{56}+d\Big(\xi_3(r) \text{Im}\ \Omega\Big)+d\Big(\xi_4(r)\text{Re}\ \Omega\Big)+\xi_5 (r) e^{8,9,10,11}.
\end{align}

With the ansatz for the metric (\ref{metrics}) and the 4-form
field strength (\ref{fourforms}), we are ready to impose the
eleven-dimensional supergravity equations of motion (\ref{sugra eq}).
The second equation in (\ref{sugra eq}), namely the Maxwell
equation for the 4-form, can be solved by,
\begin{align}\label{flux sol}
\begin{split}
\xi_1(r)=\frac {C_1} {g(r)^2}, \quad\quad\xi_2(r)=-\frac {2C_1}{g(r)^2}+\frac{C_2}{g(r)h(r)},\\
\xi_3(r)=0, \quad\quad \xi_4(r)=-\frac{3\sqrt 2\,\xi(r)}{g(r)h(r)^{1/2}},\quad\quad \xi_5(r)=0,
\end{split}
\end{align}
where the function $\xi(r)$ satisfies the differential equation,
\begin{align}\label{xi eq}
\xi ''+
   \frac{2 f' \xi '}{f}-\frac{4 h
   \left(\xi-\frac 3 2\right)}{g^2}-\frac{2 \xi
 }{h}=0.
\end{align}
From now on, we will set the dimensionful constant in (\ref{F ansatz}) to be $c= \sqrt 2$. In order for the 4-form (\ref{fourforms}) coverge to (\ref{F ansatz}) as $r\rightarrow \infty$,  one must impose $C_1 = 9\sqrt2$, $C_2 = 0$ and $\xi(\infty)= 1$.
Thus, the 4-form can be expressed as,
\begin{align}\label{F4}
F_4 = \frac {9\sqrt2} {g(r)^2}\,  J\wedge J - \frac {18\sqrt2}{g(r)^2}J\wedge e^{56}-d\left(\frac{3\sqrt 2\,\xi(r)}{g(r)h(r)^{1/2}}\text{Re}\ \Omega\right).
\end{align}

 The next step is to express the first equation in (\ref{sugra eq}),
 namely the Einstein equations, in
 our ansatz as,
\begin{align}\label{ein eq}
\begin{split}
-&\frac{g''}{2 g}-\frac{f' g'}{f g}-\frac{g'
   h'}{2 g h}-\frac{g'^2}{2 g^2}-\frac{\xi '^2}{24 g^2
   h}-\frac{\xi ^2 }{12 g^2 h^2}-\frac{h
   }{g^2}-\frac{2\left(\xi-\frac 3 2\right)^2}{3g^4}+\frac{3
    }{g}=0,\\
   -&\frac{h''}{2 h}-\frac{f' h'}{f h}-\frac{g' h'}{g h}-\frac{\xi
   '^2}{24 g^2 h}-\frac{\xi ^2  }{3 g^2
   h^2}+\frac{h
    }{g^2}+\frac{\left(\xi-\frac 3 2\right)^2
    }{3 g^4}+\frac{ 1}{h}=0,\\
   -&\frac{f''}{2 f}-\frac{f' g'}{f g}-\frac{f'
   h'}{2 f h}-\frac{f'^2}{2 f^2}+\frac{\xi
   '^2}{12 g^2 h}+\frac{\xi ^2  }{6 g^2
   h^2}+\frac{\left(\xi-\frac 3 2\right)^2 }{3 g^4}+\frac{3  }{f}=0,\\
    &\frac{8 f' g'}{f g}+\frac{4 f' h'}{f h}+\frac{3
   f'^2}{f^2}+\frac{h'^2}{2
   h^2}+\frac{4 g' h'}{g
   h}+\frac{3 g'^2}{g^2}-\frac{\xi '^2}{4 g^2
   h}\\&\qquad\qquad+\frac{\xi ^2  }{2 g^2 h^2}+\frac{2 h
   }{g^2}+\frac{\left(\xi-\frac 3 2\right)^2
   }{ g^4}-\frac{12
    }{g}-\frac{2  }{h}-\frac{12  }{f}=0,\\&\xi ''+
   \frac{2 f' \xi '}{f}-\frac{4 h
   \left(\xi-\frac 3 2\right)}{g^2}-\frac{2 \xi
    }{h}=0.
   \end{split}
   \end{align}
For our reference below, we added the Maxwell equation equation
for the 4-form (\ref{xi eq}) in the end. There are four independent functions, four Einstein equations and one Maxwell equation. Due to the Bianchi identities, only three out of four Einstein equations are independent.

As a consistency check, we can easily verify that
the Euclidian version of AdS$_5\times \mathbb{C}{\rm P}^3$,
\begin{align}\label{vac}
 f(r)=\sinh^2 r, \quad h(r)=\frac 1 2, \quad g(r)= \frac 1 2, \quad \xi(r)=1,
\end{align}
solves these equations. There is another simple solution,
\begin{align}\label{squashed}
\begin{split}
&f(r)=\frac 4 3 \left(\frac 2 3\right)^{2/3} \sinh\left(\frac 1 2 \left(\frac 3 2 \right)^{5/6} \sqrt{2} r\right)^2, \quad h(r)=\left(\frac 2 3\right)^{2/3},\\& \qquad\qquad\qquad g(r)= \frac 1 {2^{1/3} 3^{2/3}} \quad \xi(r)=\frac 4 3,
\end{split}
\end{align}
which is a stretched $\mathbb{C}{\rm P}^3$ solution \cite{Imaanpur:2012jx}.
One can see that $h(r)=2g(r)$, $i.e.$, $\mathbb{C}{\rm P}^3$ is stretched along its fiber.

\section{Boundary Conditions} \label{sec asym}

In this section, we will study boundary conditions to
instanton solutions at the infinity of AdS and at the bubble of nothing.

For  $r\rightarrow \infty$,
the solution should approach the vacuum
AdS$_5 \times \mathbb{C}{\rm P}^3$, and we can linearlize~(\ref{ein eq}). In this set of equations, three are second order differential equations for $g$, $h$, $\xi$ and one is first order
 for $f$ (modulo the redundancy by the Bianchi identities). We should also note that there is
translational invariance in $r$, which is the residual symmetry in
our gauge~(\ref{metrics}). Therefore, there are
six linearly independent modes, and they are
$e^{2(\pm \sqrt{7}-1)r}$, $e^{2(\pm\sqrt{10}-1)r}$, $e^{-2r}$, and $r \cdot e^{-2r}$. Among them, three  are normalizable and three are non-normalizable.
Note that $e^{-2r}$ and $r \cdot e^{-2r}$ are at the Breitenlohner-Freedman bound.
Conformal invariance on the boundary requires the $r \cdot e^{-2r}$ mode to vanish \cite{Klebanov:1999tb}.
This condition also guarantees that the instanton action is finite, as we will see in section \ref{inst action}.
For now, we only set
the two diverging modes,  $e^{2(\sqrt{7}-1)r}$ and
$e^{2(\sqrt{10}-1)r}$, to vanish at $r=\infty$. We will keep
the $r \cdot e^{-2r}$
mode  to be adjustable in the next couple of sections and demand it to vanish in section \ref{inst action}.

Let us turn our attention to boundary conditions at the bubble of nothing.
In order for the $S^2$ fiber to shrink to zero size,
the 4-form flux should
not have components on the $S^2$, otherwise the flux conservation
would prevent it from collapsing. Thus, $\xi(r)$ in (\ref{F4})
must be chosen in such a way that
$F_4$ is proportional to the volume form of the base $S^4$. For this
purpose, it is convenient to rewrite (\ref{F4}) as,
\begin{align}\label{F4 xi0}
\frac{1}{3\sqrt 2}\, F_4 =\frac {4 \left(\frac 3 2 -\xi(r)\right)}{g(r)^2}e^{1234}-\frac {2\,\xi(r)}{g(r)h(r)}J \wedge e^{56}+\frac {\xi'(r)}{g(r)\sqrt{h(r)}} \text{Re}\,\Omega\wedge dr .
\end{align}
Note that the second and third terms in the right-hand side
 have $h(r)$ and $\sqrt{h(r)}$ in the denominators, which should
 vanish at the bubble. However,
$e^{56}$ and $\text{Re}\ \Omega$ also go to zero since they
have the factors $h(r)$ and $\sqrt{h(r)}$ respectively.
Therefore these second and third terms vanish
 if we set $\xi=0$ and $\xi'=0$, and $F_4$ becomes proportional to
 the volume form $e^{1234}$ on the base.

Suppose the $S^2$ fiber becomes zero size at $r=r_0$.
This means we set $h(r_0)=0$. We also require $\xi(r_0) = \xi'(r_0)=0$
due our analysis in the previous paragraph. Combining these boundary
conditions with
the equations of motion (\ref{ein eq}), we find,
\begin{align}
\begin{split}\label{r=0}
\xi( r)&= \xi_0 \ (r-r_0)^2+O((r-r_0)^4),\\
f( r)&=f_0+O((r-r_0)^2),\\
g( r)&=g_0+O((r-r_0)^2),\\
h( r)&=(r-r_0)^2 + O((r-r_0)^4).
\end{split}
\end{align}
In order for the geometry to terminate smoothly, we need
$h(r) = (r-r_0)^2 + \cdots$ with the coefficient $1$
in the leading term.
This condition turns out to be implied by the Einstein equations.
This is in contrast to the case of
Witten's instanton, for which an analogue of
 $h(r) = (r-r_0)^2 + \cdots $ has to be imposed as an additional
 boundary condition.

 Thus, we find that there are three parameters $f_0$, $g_0$ and
 $\xi_0$ at the bubble. As we will see below,
 they can be fixed by demanding the three
 non-normalizable  modes,  $e^{2(\sqrt{7}-1)r}$,
$e^{2(\sqrt{10}-1)r}$, and $r \cdot e^{-2r}$, to vanish at the infinity.
The location $r_0$ of the bubble is fixed by demanding
$f(r)/\sinh^2 r \rightarrow 1$ for $r \rightarrow \infty$.

\section{Algebraic Relations} \label{anal rel}

Interestingly, both the equations of motion and the boundary conditions
defined in the last two sections are compatible with
two simple algebraic relations between the three
functions $g(r),h(r),\xi(r)$.
In fact, if we set,
\begin{align}\label{GS}
\begin{split}
g(r)&= G(h(r)),\\
\xi(r)&=S(h(r)),
\end{split}
\end{align}
and substitute them into the equations of motion (\ref{ein eq}),
we find a couple of equations that are independent of $f(r)$:
\begin{align}\label{eq S}
\begin{split}
&\frac{3}{G} -\frac{h}{G^2}-\frac {S^2}{12h^2\,G^2}-\frac{2\left(S-\frac 3 2\right)^2}{3\,G^4}+\dot G\left(\frac{S^2 }{3\, G^3 h}-\frac{1}{G}-\frac{h^2}{G^3} -\frac{h (S-\frac 3 2)^2 }{3 G^5}\right)\\&=-h'(r)^2 \left(\frac{\dot G^2}{2 G^2}-\frac{\ddot G}{2 G}+\frac{\dot G \dot S^2}{24
   G^3}-\frac{\dot G}{2 h G}-\frac{\dot S^2}{24 h
   G^2}\right)
\end{split}
\end{align}
and
\begin{align}\label{eq G}
\begin{split}
&-\frac{4h \left(S-\frac 3 2\right)}{G^2}-\frac{2S}{h}+\dot S\left( 2+\frac{2h^2}{G^2}-\frac{2 S^2}{3hG^2}+\frac{2h\left(S-\frac 3 2\right)^2}{3G^4}\right)\\&=-h'(r)^2 \left(\ddot S-\frac{2 \dot G \dot S}{G}-\frac{\dot S^3}{12
   G^2}\right),
\end{split}
\end{align}
where $\dot {} = d/dh$.
Demanding that these equations hold independently of $h'(r)$,  we obtain four
differential equations on $G(h)$ and $S(h)$ with respect to $h$.
Remarkably, these four equations can be solved algebraically by imposing the simple relations,
\begin{align}
\begin{split}\label{relations}
\xi&=3-\sqrt {2g}\left(3g+h\right),\\
1&=\sqrt{2g}\big(h+g\big).
\end{split}
\end{align}

These algebraic relations are also consistent with the boundary
conditions at $r=r_0$ and $\infty$: Setting $h(r) = (r-r_0)^2$
gives $g(r) = 2^{-1/3}$ and $\xi(r) = 2^{4/3}(r-r_0)^2$ as expected
at the bubble, and
$h = 1/2$ at $r=\infty$
gives $g=1/2$ and $\xi=1$ as required for AdS$_5 \times \mathbb{C}{\rm P}^3$.
We found these relations experimentally, and it would be
interesting to find their deeper origins. In the following, we will use them
to numerically integrate the rest of the equations of motion.

\section{Numerical solution} \label{num sol}
Using of the algebraic relations (\ref{relations}), the equations of
motion (\ref{ein eq}) collapse to the following
three equations for
the two functions $f(r)$ and $g(r)$,
\begin{align}
\begin{split}
-&\frac{g''}{2g}-\frac{g'^2}{4g^2}-\frac{f'g'}{fg}-\frac{1}{6g^4}(1-5\sqrt 2 g^{3/2}+12g^3)=0,\\
-&\frac{f''}{2f}-\frac{f'^2}{2f^2}
 -\frac{3f'g'}{4fg}\frac{1-(2g)^{3/2}}{1-\sqrt{2} g^{3/2}}+\frac{3g'^2}{2g^2} \frac{\sqrt{2}g^{3/2}}{1-\sqrt{2} g^{3/2}}+\frac{1}{12g^4}(1-8\sqrt{2}g^{3/2}+48g^3)+\frac{3}{f}=0,\\&
\frac{3f'^2}{f^2}+\frac{6 f'g'}{fg}\frac{1-(2g)^{3/2}}{1-\sqrt{2} g^{3/2}}+\frac{g'^2}{8g^2}\frac{(9-96\sqrt 2 g^{3/2}+192 g^3)}{(1-\sqrt2 g^{3/2})^2}+\frac{1}{4g^4} \frac{1-5\sqrt 2 g^{3/2}}{1-\sqrt2 g^{3/2}}-\frac{12}{f}=0.
\end{split}
\end{align}
Only two of these three equations are independent. Eliminating $f(r)$,
one finds the following equation for $g(r)$,
\begin{align}\label{g eq}
\begin{split}&
\left(1-\sqrt{2} g^{3/2}\right)^2 g^6 \left(4 g''' g'-5
   g''^2\right)-\frac 3 4 \left(1+14
   \sqrt{2} g^{3/2}-42 g^3\right) g^4 g'^4\\-&
   \left(5-16 \sqrt{2} g^{3/2}+22 g^3\right) g^5 g'^2 g''-2 \left(1-3 \sqrt{2} g^{3/2}\right) \left(1-2 \sqrt{2} g^{3/2}\right) \left(1-\sqrt{2} g^{3/2}\right)^2 g^3 g''\\-& \left(1-\sqrt{2} g^{3/2}\right)^2 \left(4-9 \sqrt{2} g^{3/2}-12 g^3\right) g^2 g'^2\\-&\frac 1 9 \left(1-3 \sqrt{2} g^{3/2}\right)^2 \left(1-2 \sqrt{2} g^{3/2}\right)^2 \left(1-\sqrt{2} g^{3/2}\right)^2=0.
   \end{split}
\end{align}
This is a third order differential equation for $g(r)$. Since it does
not depend on $r$ explicitly, one can lower its order by one. The numerical integration of this equation shows the desired behavior, $i.e.$, $g(r)$ goes from $2^{-1/3}$ to $1/2$
as $r$ goes from $r_0$ to $\infty$.

\begin{figure}[h!]
 \centering
  \includegraphics[width=0.9\textwidth]{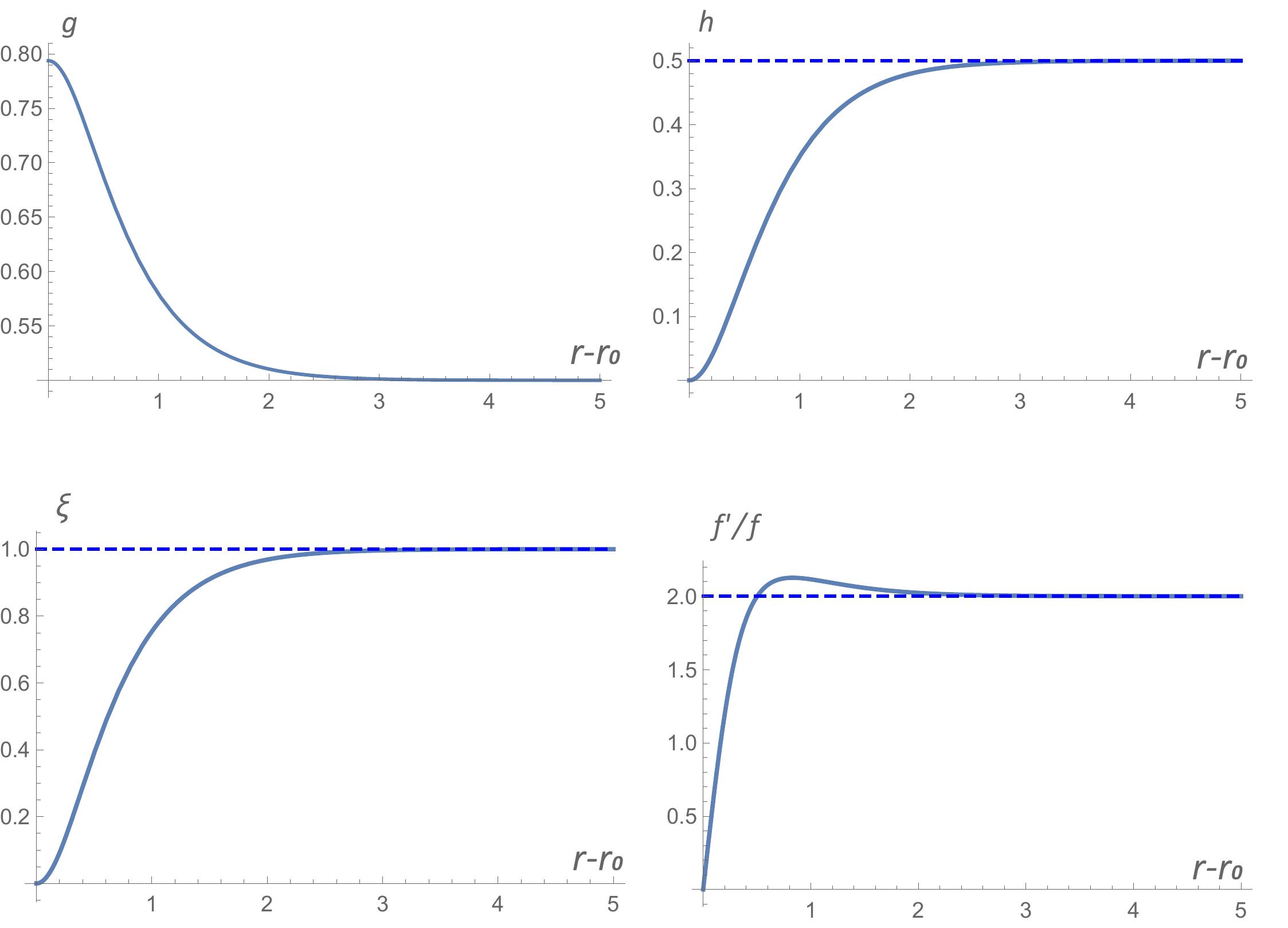}
 \caption{Numerical solution to the equation (\ref{g eq})}
 \end{figure}

We would like to comment on two important features of our
numerical solutions:
\begin{enumerate}
\item
The equation for $g(r)$ is singular at the bubble,
and the coeffieint of $g'''$ vanishes with $g(r_0)= 2^{-1/3}$.
Therefore, instead of numerically integrating the equation with
the initial value of  $g(r_0)= 2^{-1/3}$, we computed first few terms
in the Taylor expansion analytically and matched them to
a numerical solution.
\item In this section, we are not requiring the $r \cdot e^{-2r}$ mode to vanish at
$r \rightarrow \infty$. Thus, we are left with one free parameter $f_0$, which is the size of the bubble.
\end{enumerate}

We presented the typical behavior of the solution on the Figure 1,
where we set $f_0=1$. The horizontal axis is $r-r_0$, where
$r_0$ is fixed by demanding $f(r) / \sinh^2 r \rightarrow 1$
as $r \rightarrow \infty$.
One can see that the solution exhibits the desired behavior: $h(r)\rightarrow 1 /2$, $g(r)\rightarrow 1 /2$, $\xi(r)\rightarrow 0$ and $ f'(r)/f(r) \rightarrow 2$ as $r\rightarrow \infty$.

\section{Instanton Action}\label{inst action}

We have shown that there is a family of solutions
parameterized by the size $f_0$ of the bubble,
which approach AdS$_5\times \mathbb{C}{\rm P}^3$ at infinity.
However, there is one more non-normalizable mode
$r \cdot e^{-2r}$ we need to fix. In this section,
we show that one can  turn off this mode by adjusting~$f_0$.
This also makes the instanton action finite.

For general value of the parameter $f_0$, the solution at the infinity behaves as
\begin{align}\label{g asym}
g(r) = \frac 1 2 + \big(a(f_0) + b(f_0) r \big)e^{-2r}+\cdots.
\end{align}
Numerically, $b(f_0)$ vanishes at $f_0=0.6203025..$.
To show that $b(f_0)$ can be set exactly equal to zero by adjusting $f_0$, we
present Figure 2 for
$\alpha(r) = \left(g(r) -\frac 1 2\right)e^{2r}$. Since $\alpha(r) \sim a(f_0) + b(f_0) r$ for $r \rightarrow \infty$,
we see that $b(f_0)$ changes
its sign near $0.62$. Therefore, there must be $f_0$
near $0.62$ such that $b(f_0)=0$.

\begin{figure}[h!]
 \centering
  \includegraphics[width=0.9\textwidth]{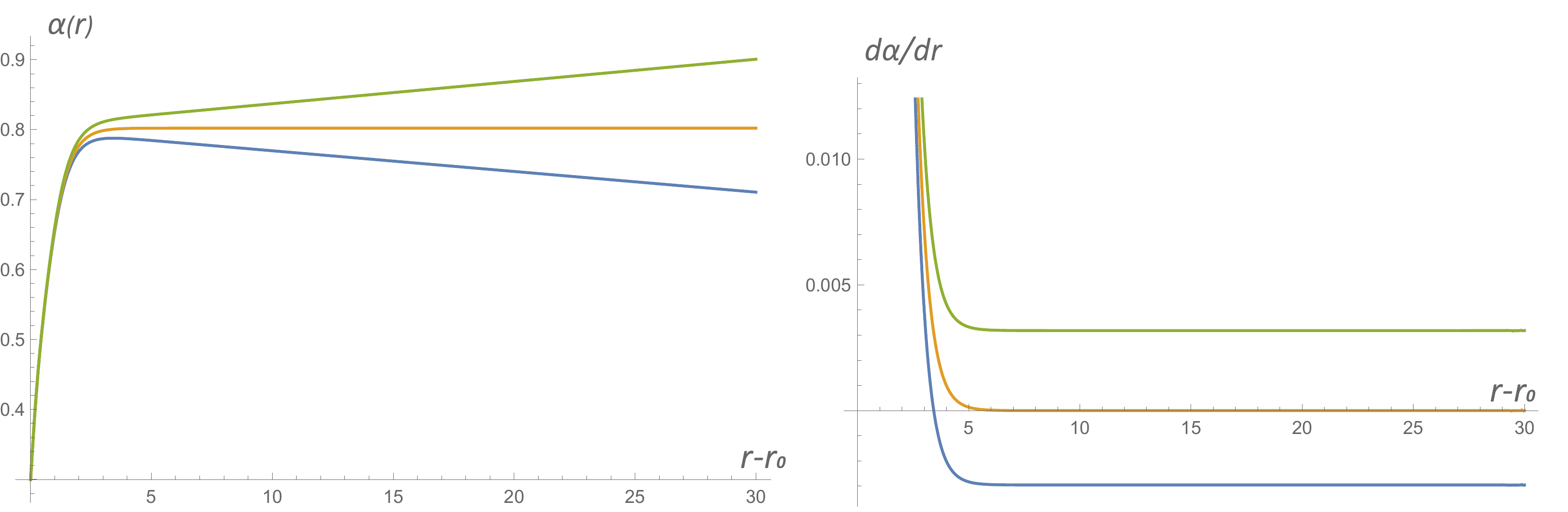}
 \caption{The graphs of $\alpha(r)$ and its derivative. The
 horizontal axises are $r-r_0$. Values of $f_0$ from top to bottom are 0.63, 0.62, 0.61 respectively.}
 \end{figure}

The action for 11-dimensional supergravity takes the form,
\begin{align}
S = \int_{M_{11}}  \sqrt{\det G} \left( \frac 1 4 R - \frac 1 {48} F_{MNPQ}F^{MNPQ}+\dots\right),
\end{align}
where we have ignored the Chern-Simons terms and fermions,
which are irrelevant to our discussion.
Using the supergravity equations (\ref{sugra eq}), the 4-form kinetic
term is related to the Einstein term, $F_{MNPQ}F^{MNPQ}=36 R$. Therefore,
the instanton action reduces to,
\begin{align}
S = -\frac 1 2 \int_{M_{11}} \sqrt{\det G} \,\, R.
\end{align}
It is straightforward to see that,
with the $r \cdot e^{-2r}$ mode removed,
the instanton action is finite and positive after subtracting the value for the vacuum AdS.

We conclude that
AdS$_5\times \mathbb{C}{\rm P}^3$
is unstable due to the finite action instanton.

\section{Instability of AdS$_{\bf 5} \times  \frac{{\bf SU(3)}}{{\bf U(1)}
\times {\bf U(1)}}$ }\label{cp1 cp2 inst}

In this section, we will show that AdS$_5\times \frac {SU(3)}{U(1)\times U(1)}$ model has an instanton that mediates its decay.
It is not know whether this solution is perturbative stable or not. In either case,
the existence of the bubble of nothing solution shown here means that it is unstable.

To construct the solution, let us first review the geometry of $\frac {SU(3)}{U(1)\times U(1)}$. It can be viewed as a
 flag manifold $\mathbb F(1,2,3)$ or a twistor space over
 $\mathbb{C}{\rm P}^2$. It admits K\"{a}hler structure with an Einstein metric
 and therefore is a solution of the supergravity equations of motion.  It is
 also an $S^2$ fibration over $\mathbb{C}{\rm P}^2$ base.
 The last fact is best understood from the coset point of view.
 One can choose $SU(2)$ subgroup in $SU(3)$ and decompose $SU(3)= \frac {SU(3)}{SU(2)}\times SU(2)$. The former term in the product is homogeneous space  $S^5$ and the latter is $S^3$. Therefore, $SU(3)$ is a $S^3$ fibration over $S^5$. The two $U(1)$ subgroups in the denominator of the coset $\frac{SU(3)}{U(1)\times U(1)}$ turn each sphere into complex projective space resulting in $S^2\hookrightarrow \frac{SU(3)}{U(1)\times U(1)}\rightarrow \mathbb{C}{\rm P}^2$ fibration.

 It worth mentioning that both $\mathbb{C}{\rm P}^3$ and $\frac{SU(3)}{U(1)\times U(1)}$ are twistor spaces of $S^4$ and $\mathbb{C}{\rm P}^3$ \cite{Tomasiello:2007eq}. This fact seems to be the main reason of the similarity between the collapsing solutions of the models.
 Choosing the vielbein as \cite{Duff:1986hr,Page:1984ac}:
\begin{align}
\begin{split}
e_1&=\sqrt{2g(r)}d\mu,\\
e_2&=\sqrt{\frac {g(r)}2}\sin\mu\,\Sigma_{1},\\
e_3&=\sqrt{\frac {g(r)}2}\sin\mu\,\Sigma_{2},\\
e_4&=\sqrt{\frac {g(r)}2}\sin\mu\cos\mu\,\Sigma_{3},\\
e_5&= \sqrt{h(r)} (d\theta - A_1\sin \phi +A_2\cos \phi ),\\
e_6&=\sqrt{h(r)}\sin \theta(d\phi - \cot\theta(A_1\cos\phi +A_2\sin\phi )+A_3),
\end{split}
\end{align}
with
\begin{align}
\begin{split}
\Sigma_1&= \cos\gamma\,d\alpha+\sin\gamma \,d\beta,\\
\Sigma_2&= -\sin\gamma\,d\alpha+\cos\gamma \,d\beta,\\
\Sigma_3&= d\gamma+\cos\alpha \, d\beta,\\
A_1&=\cos\mu\,\Sigma_{1},\\
A_2&=\cos\mu\,\Sigma_{2},\\
A_3&=\frac 1 2 \left(1+\cos^2\mu\right)\,\Sigma_{3},
\end{split}
\end{align}
 all the formulas and results become exactly the same as in $\mathbb{C}{\rm P}^3$ case.  Namely,  $\frac{SU(3)}{U(1)\times U(1)}$  has the $SU(3)$-structure defined by (\ref{su3 def}) and has torsion classes~(\ref{torsion}). Since the $SU(3)$-structure is the same, the ansatz for the flux will have the same solution~(\ref{flux sol}). Finally and most importantly, the equations of motion of supergravity take exactly the same form~ (\ref{ein eq}). The last fact makes all the results of the previous sections applicable to this case. The only thing which is different is the expression of the vielbein in terms of the coordinates.

One might wonder why the equations are exactly the same. It follows
from the fact that we constructed the ansatz which respects all the symmetries of the model with squashed fiber. The bases of the compact manifold in both cases are Einstein manifolds and therefore their contribution to the Einstein equations will enter in a similar manner. Besides, the $SU(3)$-structure is rooted in the twistor origin of both spaces and it is used to build the ansatz for the flux. Because of the same origin, it gives the same result in both cases.

\section{Discussion}\label{discussion}

We would like to end this paper by explaining how our solution
evades the issue raised in \cite{Young:1984jv}.
Suppose we try to collapse a $d$-dimensional sphere in the internal space
supported by a flux, at the bubble located at $r=r_0$.
The flux gives contribution to the Einstein equations proportional to $1/h(r)^d$, where $h(r)$ is the square radius of the sphere, while the contribution from curvature is proportional to $1/(r-r_0)^2$. Taking into account that $h(r)\sim (r-r_0)^2$ for the smoothness,
it was argued in \cite{Young:1984jv} that
 the only possible way to make this two terms of the same order is to set $d=1$, $i.e.$, a circle. However, this
  does not apply to our case since the amount of flux on the sphere can vary.

   It is instructive to see it explicitly in the Einstein equation (\ref{ein eq}) for $h(r)$,
\begin{align}
-\frac{h''}{2 h}-\frac{f' h'}{f h}-\frac{g' h'}{g h}-\frac{\xi
   '^2}{24 g^2 h}+\frac{h
   }{g^2}+\frac{\left(\xi-\frac 3 2\right)^2
    }{3 g^4}+\frac{1 }{h}-\frac{\xi ^2  }{3 g^2
   h^2}=0.
\end{align}
 One can see that curvature contribution (first term) is of order $1/(r-r_0)^2$, while the flux (last term) is of order $\xi(r) ^2/h(r)^2$ (flux for the $S^4$ is proportional to the 4th power of $g(r)$ as it should). This is consistent with the estimate of \cite{Young:1984jv} for $d=2$.
 However,
 in our solution, these two terms can balance each other since $\xi(r)\rightarrow 0$ when $h(r)\rightarrow0$.
 In this way, the flux evaporates from the $S^2$ fiber, and the Einstein equations can be satisfied.

Another possibility to deal with flux conservation is
to introduce a domain wall at the bubble to absorb the flux. 
This idea was used in \cite{Horowitz:2007pr} to collapse
the supersymmetry breaking $S^1$ in
the AdS$_5 \times S^5/\Gamma$ geometry of  \cite{Kachru:1998ys}.
More recently, instanton solutions with $S^2$ collapsing have been
constructed in some models in six dimensions.
 Blanco-Pillardo, {\it et al.} \cite{Blanco} considered the Einstein gravity in 
 six dimensions coupled to $SU(2)$ Yang-Mills gauge field and an 
 adjoint Higgs field with a specific potential 
to break the gauge group to $U(1)$ and found a smooth solution of this type.
Brown and Dahlen
\cite{Brown} considred the Einstein-Maxwell system without the Higgs and added a domain wall as a source.
It would be interesting to realize such solutions in a low energy effective theory of M/string theory in a controlled approximation.

According to \cite{Martin:2008pf}, AdS$_5 \times \mathbb{C}{\rm P}^1\times \mathbb{C}{\rm P}^2$ and AdS$_5 \times \mathbb{C}{\rm P}^1\times \mathbb{C}{\rm P}^1\times \mathbb{C}{\rm P}^1$ are not stable perturbatively.
Thus, we do not need instantons  for these geometries to be consistent with
the conjecture of \cite{Ooguri:2016pdq}.
In fact, our ansatz is not applicable to them since
the configurations are too restrictive for the fluxes to slip off. In our solution, it is important that $\mathbb{C}{\rm P}^3$ has the
non-trivial fibration structure and not a direct product since this allows our non-trivial solution to the 4-form equations.
A similar argument applies to the Freud-Rubin type compactification (when the flux is proportional to the volume form of the AdS) and the compactifications where the flux is proportional to the volume form of the compact space (unless one can turn on other lower dimensional form).

Finally, we want to mention that it is possible that our solution is related to resolved M-theory conifold solutions with $G_2$-holonomy\cite{Gibbons:1989er}. They are Ricci-flat seven-dimensional manifolds which have conic structure and its metrics reads,
\begin{align}
ds^2= \frac{1}{1-\frac1 {r^4}}dr^2 + \frac 1 4 r^2\left(1-\frac1 {r^4}\right) |d_A u|^2+ \frac{r^2} 2 ds^2_{M_4},
\end{align}
where $M_4$ is either $S^4$ or $\mathbb{C}{\rm P}^2$ and $|d_A u|^2$ is metrics on the $S^2$ fiber which are exactly the same as in the present paper for $\mathbb{C}{\rm P}^3$ and $\frac {SU(3)}{U(1)\times U(1)}$ respectively. Moreover, the $S^2$ fiber collapse at finite $r=1$,
while $M_4$ radius stays finite. Unfortunately, the radii grow linearly at infinity.  It may be possible to construct a desirable solution by multiplying this geometry with the flat $\mathbb{R}^4$ and by adding some flux along the conifold in order to change the behavior at infinity from the linear growth to constant.  An idea along this line may allow us to generalize our solutions further.

\section*{Acknowledgement}

We would like to thank G.~Horowitz, S.~Gukov, I.~Klebanov, and H.~Reall for discussions.
This research is supported in part by
U.S.\ Department of Energy grant DE-SC0011632.
H.O. is also supported in part
 by the Simons Investigator Award,
by the World Premier International Research Center Initiative,
MEXT, Japan,
by JSPS Grant-in-Aid for Scientific Research C-26400240,
and by JSPS Grant-in-Aid for Scientific Research on Innovative Areas
15H05895.

\end{document}